\RequirePackage{fix-cm}
\documentclass[smallextended]{svjour3}       
\smartqed  
\usepackage{graphicx,ulem}
\usepackage{cite}
\usepackage{graphicx,amsmath,amssymb,latexsym}
\usepackage{color}
\usepackage{epstopdf}
\usepackage{graphics}
\usepackage{url}
\usepackage{bbm}

\usepackage{subfigure}

\DeclareMathOperator*{\argmin}{arg\!\min}

\begin{document}

\title{Strategic Arrivals to Queues Offering Priority Service}


\author{Rajat Talak        \and
        D. Manjunath \and
        Alexandre Proutiere 
}

\authorrunning{Talak, Manjunath and Proutiere}

%


\maketitle


\begin{abstract}
We consider strategic arrivals to a FCFS service system that starts
service at a fixed time and has to serve a fixed number of customers,
e.g., an airplane boarding system. Arriving early induces a higher
waiting cost (waiting before service begins) while arriving late
induces a cost because earlier arrivals take the better seats. We
first consider arrivals of heterogenous customers that choose arrival
times to minimize the weighted sum of waiting cost and and cost due to
expected number of predecessors. We characterize the unique Nash
equilibria for this system.

Next, we consider a system offering $L$ levels of priority service
with a FCFS queue for each priority level.  Higher priorties are
charged higher admission prices. Customers make two choices---time of
arrival and priority of service.  We show that the Nash equilibrium
corresponds to the customer types being divided into $L$ intervals and
customers belonging to each interval choosing the same priority
level. We further analyze the net revenue to the server and consider
revenue maximising strategies---number of priority levels and
pricing. Numerical results show that with only a small number
  of queues (two or three) the server can obtain nearly the maximum
revenue.
\end{abstract}

%

\section{Introduction}
\label{sec:intro}

Consider an airline that starts boarding a plane at time $0$ and the
$N$ customers that are booked to fly on the plane all arrive before
boarding starts and wait in a FCFS queue. A customer that has a better
rank in the queue gets a better choice of seats and luggage space.
 However, a better queue rank is achieved by arriving early and
  hence incurring a cost due to the waiting time before service
  (boarding) begins. Thus, with each customer we can associate a total
  inconvenience cost that has two components---(1)~the \textit{waiting
    cost} from the inconvenience due to arriving early (before
  boarding time), and (2)~the \textit{boarding cost} from the
  inconvenience due to customers that are served before the focal
  customer, i.e., inconvenience due to its rank or position in the
  boarding queue. An immediate model would be to assume that these
  costs are separable, and that the total cost is a weighted sum of
  the waiting time and the rank in the boarding sequence. (Note that
  the second component accounts for possibly non negligible service
  time once boarding begins.) Each customer then has to choose the
  time at which to arrive into the queue, i.e., the waiting time
  before service begins. Of course, for a given customer the queue
  rank that it obtains depends on the choice of arriving (and hence
  waiting) times of the other customers.  In this paper we assume that
  the customers are strategic and that they choose their arrival time
  to minimize their individual cost. We also assume that none of the
  customers renege and that all $N$ take service. Thus we have the
  rudiments of the \textit{airplane boarding game} which we will
  define more generally in the following.

Now assume that the customers are from a heterogeneous population and
that the different types of customers calculate their total costs
differently. Specifically, for the linear cost function described in
the preceding, different types of customers have different weights for
the queue rank and for the waiting time before boarding. In this case,
rather than have a single queue and hence provide egalitarian service,
the airline could reduce the social cost, and possibly increase its
revenue, by providing service differentiation as follows. Maintain two
FCFS queues with queue $2$ having strict boarding priority over queue
$1,$ i.e., board customers from queue $2$ before boarding customers
from queue $1.$ As before, assume that all the customers
  arrive before boarding begins and that there is no reneging by any
  of the $N$ customers.  The airline could now charge a fixed premium
to the customers who join the higher priority queue. To obtain better
seats, i.e., to make their boarding costs lower, the customers can now
choose to join the high priority queue and trade off some waiting time
(before boarding begins) for payment of the premium in the
higher priority queue. In such a system the customers make two
choices---the arrival time and the queue that they plan to join and
hence the premium that they will be paying. Once again, the actual
total cost for each customer depends on the choice of these parameters
by the other customers. We thus have the outline of the basic
\textit{airplane boarding game} in which customers strategically choose
two parameters to minimize their individual total cost.

Several budget airlines in fact follow a simple version of the system
described above and motivates this paper. In these airlines, seat
numbers are not issued at check-in and typically two FCFS queues are
maintained at the boarding gate. The premium queue allows service
differentiation to a heterogeneous population where different types of
customers have different relative values for their time and money and
provides an opportunity for the airline to make additional revenue.
If there is significant heterogeneity in the customer population then
increased differentiation can provide increased revenue. Thus, rather
than provide just two priorities, the airline could provide $L$
priority queues with customers in priority $l$ having strict priority
over those of strictly less than $l.$ In this paper we consider this
general case of the following \textit{airplane boarding game.}

$N$ customers are to be served by a single server starting service at
time $0.$ Different types of customers will be served through $L$ FCFS
priority queues with higher priorities charging higher
premiums. Associated with each type of customer is a total cost
function that depends on the waiting time, the rank in the boarding
sequence, and the premium paid to receive service from the queue of
choice. The distribution of the types in the customer population is
assumed known to the server and to the other customers. Each customer
strategically chooses two parameters, the arrival time and the
priority level of the queue from which it wants to receive service, to
minimize its expected total cost. None of the $N$ customers
  renege and all of them obtain service and this is known to all the
  customers.

\subsection{Preview of the Results}

In this paper our objectives are twofold: (1)~studying customers'
strategic behavior in their choice of the priority and of the arrival
time, and~(2)~analyzing how the server may tune the service
parameters, the number of priority levels and the charge for each of
these priority levels, to maximize its revenue. The following is the
summary of the key results in the paper.

\begin{enumerate}
\item In Section~\ref{sec:single-queue}, we consider a system where
  customers choose a single priority parameter that we call the grade
  of service. The service provider offers a continuum of grades of
  service and defines a pricing function that maps the service grade
  to a price. Higher service grades have strictly higher priority in
  boarding. Customers are strategic and choose the service grade and
  hence the price. For this case we show that the revenue is
  independent of the pricing function.

  Interestingly, we will see that the model considered in this section
  is very general in that it is applicable to `finite duration' games
  like in the airline boarding game where a finite number of customers
  are to be allocated a finite resource. For this case our results
  should be interpreted in the spirit of \cite{Gueant10}. The results
  are also applicable to `infinite duration' games where the customer
  arrivals and service processes are in a stationary regime; in this
  case the results are to be interpreted in the spirit of, e.g.,
  \cite{Lui85,Glazer86,Doroudi13}.

\item In Section~\ref{sec:multiple-queues}, we consider the system
  where the server maintains several FCFS queues at different priority
  levels. Customers pay a higher price to join a queue with higher
  priority. The number of priority levels and the price of service
  from these priority levels is assumed known to all the customers. In
  this case the customers have to select two parameters---the arrival
  time before boarding starts, and the priority of the queue from
  which they will receive service. We show that when the prices are
  fixed, the game admits a unique Nash equilibrium that we
  characterize exactly. We also numerically illustrate the strategic
  choices of the priority parameters at equilibrium.

\item Finally, in Section~\ref{sec:revenue}, we investigate how the
  service provider may set the prices to join the various queues to
  maximize its revenue. When the number of priority queues are fixed,
  we show that these prices can be computed by solving a simple
  dynamic program. Numerically, we observe that the revenue increases
  with the number of available queues. However, the marginal gain of
  operating additional queues vanishes rapidly with increasing number
  of priorities. In fact we see that for several examples, maintaining
  a small number number of queues (two or three) may actually yield a
  revenue that is almost optimal, i.e., two or three levels priority
  levels is enough to extract a `close to maximum' revenue.
\end{enumerate}

All proofs are carried in the appendix.

In the next section we provide a brief overview of the relevant
literature and delineate them from the results of this paper.

\section{Related Work}
\label{sec:literature}

The \textit{meeting game} and the concert queueing game have
similarities to the airplane boarding game . We first describe these
briefly.

A meeting is scheduled to start at time $t$ but the participants
arrive at random times and it usually starts at a random time $T$ when
quorum is achieved. Knowing that the meeting does not necessarily
start at time $t,$ participants $i$ chooses to arrive at $\tau_i$ to
minimize her costs due to waiting (a function of $(T-\tau_i)^{+}$),
due to inconvenience (a function of $(\tau_i -T)^{+}$), and due to
loss of reputation (a function of $(\tau_i - t)^{+}$). The knowledge
of the structure and distribution of these costs among the
participants can be strategically used by participant $i$ to choose
$\tau_i.$ This meeting game has been introduced and studied in
\cite{Gueant10}. A variant of the meeting game is the `concert
queueing game'. Here the concert hall opens at time $t$ and the
attendees queue up for FCFS service; each attendee requires a random
service time. Arriving late has the drawback that the better seats
will be taken by those who came earlier, whereas arriving early
induces a cost of waiting. Attendees strategically choose their
arrival time distribution to mitigate these opposing costs. Several
versions of this game have been studied in
\cite{Jain11,Juneja12a,Juneja12b}.

In the airplane boarding game all customers arrive before the start of
service which is unlike in the meeting game and in the concert
queueing game. Further, all analyses on the concert queueing game
consider a single queue while we consider multiple priority queues
providing differentiated service to a heterogeneous customer
population that values time, money and queue rank (boarding costs)
differently.  A finite number of customers being served in a
  finite time is also considered in \cite{Lariviere04}. They consider
  a discrete time system in which the customers incur a congestion
  cost due to other customers arriving in the same slot. This
  congestion cost could also be time-dependent. The airline boarding
  game that we describe here does not consider such a congestion
  cost.

Queueing systems with customers who value delays differently have been
investigated in the queueing and economics literature. Specifically,
providing lower delays for customers with higher delay costs and
extracting a commensurate payment from them has been of interest for a
while now. In an early work on pricing based on differential service,
\cite{Kleinrock67} considered a queue in which customers bribe the
server and the server provides a `highest bribe first'
service. Strategic customers in such a queue has been considered in
\cite{Balachandran72,Lui85,Glazer86}. In fact, our
Theorem~\ref{thm:ne_single_p} in Section~\ref{sec:single-queue} is a
generalization of the equilibrium analysis in \cite{Lui85,Glazer86}.
Providing differentiated service to a heterogeneous population through
priority queues and pricing the priorities has also been considered
in, among others, \cite{Marchand74,Rao98}. More recent work on
priority pricing is in \cite{Afeche04,Kittsteiner05} where only the
expected delay contributes to the cost and the weights have a specific
form. Priorities may also be auctioned like in \cite{Lui85,Glazer86}.

The models in all of the preceding studies have considered what can be
termed a `simple' system where there is only one parameter that the
customer chooses. It is the arrival time in \cite{Gueant10}, the bid
value in \cite{Kleinrock67,Balachandran72,Lui85,Glazer86} and the
priority value in \cite{Marchand74,Rao98,Afeche04,Kittsteiner05}.  In
contrast, in our model, the customer has to choose two
parameters---the waiting time \textit{and} the priority level. The
closest system to the one that we consider is that in
\cite{Honnappa11} where the attendees choose both the arrival time and
the queue to join. The key difference with our model is that in
\cite{Honnappa11} the queues start their service at different times
and could be serving in parallel. Thus in this case there is
`isolation' between the customers that choose different
queues.

A more recent work is that of \cite{Doroudi13} which considers a
priority queueing system with a pricing menu. Here the customers have
to declare their delay valuation and the focus is on deriving
incentive compatible revenue maximizing pricing functions. In our
model, type is private information and does not have to be revealed
and the choice of the grade of service is made by the customer and not
by the service provider.

\section{Single Queue: Choosing the Arrival Time}
\label{sec:single-queue}

A plane is to board $N$ customers starting at time $0.$ The $N$
customers arrive at different times before the boarding starts and
form a FCFS queue. The population from which the $N$ customers are
drawn is heterogeneous in that different customers value their waiting
time differently.  Let $\mathcal{V} = [A, B] \subset \mathbb{R}^{+}$
be the set of all customer types, and let it be endowed with a
probability measure that has a cumulative distribution function (CDF)
given by $\mathcal{G}.$ We assume $\mathcal{G}$ to be a continuous
distribution.

A customer has to decide how much before the boarding time she has to
arrive; we will denote this choice of waiting time (before
  boarding begins) by $T(v),$ for every $v \in \mathcal{V}.$
Therefore, waiting time of $T(v)$ implies that the customer arrives at
time $-T(v)$.  We assume that all customers of the same type make the
same decision. Further, we assume that none of the $N$
  customers renege and all of them join the queue before time $t=0.$
If a customer of class $v \in \mathcal{V}$ arrives at time $-t$ then
her cost is given by
\begin{equation}
  \label{eq:cost}
  c_v(t) = N F(t) + n_v g(t)
\end{equation}
where $F(t)$ denotes the fraction of customers that choose to arrive
before $-t$ on average, $g(t)$ is the cost of waiting for $t$ time
units, and $n_v$ is the weight that a type $v$ customer assigns to
it's waiting time cost. Notice that $F(t)$ is a monotonically
decreasing function. Further, $g(t)$, being the cost of waiting for
time $t$, is a monotonically strictly increasing function in $t$. We
also assume that $g$ is continuous and differentiable function with
$g(0) = 0.$ Note that $g(\cdot)$ is independent of the customer type
$v$. Without loss of generality we let $n_v$ to be a monotonically
strictly increasing function in $v.$ Thus $NF(t)$ represents boarding
cost, or the cost corresponding to the customer's position in the
queue. This cost includes the inconvenience due to the
  position in the boarding sequence and the service time of customers
  ahead of the focal customer after boarding begins. $n_v g(t)$ is
the total waiting cost for a customer of type $v.$ Now, $F(t)$ is
given by
\begin{equation}
  \label{eq:def_ft}
  F(t) = \int \mathbbm{1}_{T(v) \geq t} \ d\mathcal{G}(v).
\end{equation}

An arriving customer of type $v$ chooses $T(v)$ to minimize its
expected cost where the expectation is taken over the customer type
distribution $\mathcal{G}$. We seek an arrival profile $v \mapsto
T(v)$ that minimizes the individual expected cost for all
customers. This is also the Nash equilibrium (NE) policy as defined
below.
\begin{definition}
  $T^{NE}:\mathcal{V} \rightarrow [0, +\infty)$ is a Nash equilibrium
    policy if
  \begin{equation}
    \label{eq:ne_def_single_p}
    T^{\text{NE}}(v) = \argmin_{t \geq 0} c_{v}(t),
  \end{equation}
  for every $v \in \mathcal{V}.$ \footnote{This definition of NE is
  consistent with that found in the literature on non-atomic
  games~\cite{AliKhan97}.}
\end{definition}

Since customers with larger $v$ value their time more than those with
smaller $v,$ we would expect that customers with larger $v$ should
arrive later in a NE policy. We prove this in
Lemma~\ref{lem:ne_property_single_p}. Further, in
Theorem~\ref{thm:ne_single_p} we show that there is a unique NE and
also characterize this policy.

\begin{lemma}
  \label{lem:ne_property_single_p}
  Assume that a NE policy $T^{NE}$ exists. Then $T^{\text{NE}}(v)$ is
  a non-increasing function in $v.$
\end{lemma}
\proof{See Appendix~\ref{app:pf:lem:ne_property_single_p}. \hfill
  \qed}

\begin{theorem}
  \label{thm:ne_single_p}
  If $g(\cdot)$ is such that there exists a NE policy then it is
  unique and is given by
  \begin{equation}
  T^{\text{NE}}(v) = g^{-1}\left( \int_{v}^{B}N \frac{d\mathcal{G}(x)}{n_x}\right).
  \end{equation}
\end{theorem}

\proof{See Appendix~\ref{app:pf:thm:ne_single_p}. \hfill \qed}

\begin{remark}
  In this paper we will focus on the characterisation of the NE when
  it exists. The question of the existence of the NE is answered in
  \cite{AliKhan97, Matyszkiel2000, MasColell84} where it is shown that
  in non-atomic games that have a continuum of players, there exists a
  pure strategy NE and it is unique. For our model, i.e. considered
  assumptions on $g$ and $n_v$, the NE policy exists. More general
  conditions on $g(\cdot)$ and $n_{v}$ that yield a NE can be derived
  from~\cite{Matyszkiel2000}.
\end{remark}

Consider an example system where $g(t) = t^r$ for any $r > 0.$ For
this system, the NE policy exists and, as proved, is unique. Thus we
see that there is a non empty set of $g(\cdot)$ for which
Theorem~\ref{thm:ne_single_p} holds.


\subsection{Extensions and Generalizations}
\label{sec:single-generalizations}

Lemma~\ref{lem:ne_property_single_p} and Theorem~\ref{thm:ne_single_p}
can be extended as follows.

\begin{enumerate}
  \item The preceding results will follow even for the following total
    cost function for a customer of type $v.$ Let $h(x)$ be an
    increasing and continuously differentiable function in $x$ and let
    the cost for a customer of type $v,$ when she arrives $t$ units
    of time before boarding, be
    \begin{equation}
      \label{eq:local_1}
      c_{v}(t) = h\left( F(t) \right) + n_v g(t).
    \end{equation}
    Note that $h(\cdot)$ is independent of the customer type $v$.
    Even for this case Lemma~\ref{lem:ne_property_single_p} and
    Theorem~\ref{thm:ne_single_p} hold except that the unique NE
    arrival profile would be given by
    \begin{equation}
      T^{\text{NE}}(v) = g^{-1}\left(
      \int_{v}^{B}\frac{h'(\mathcal{G}(x))}{n_x}~d\mathcal{G}(x)\right),
    \end{equation}
    where $h^\prime(x)$ denotes derivative of $h.$

  \item In~\eqref{eq:local_1}, note that $h(F(t))$ is a specific
    decreasing function of $t$; as $h(\cdot)$ and $F(\cdot)$ are
    increasing and decreasing functions, respectively. Instead, we
    could also replace $t$ by any decreasing function of $t.$ This is
    true because if $g_{1}$ and $g_{2}$ are two decreasing functions
    over the same domain $\mathcal{D},$ then we can find an increasing
    function $h$ such that $g_{1}(x) = h(g_{2}(x))$ for all $x \in
    \mathcal{D}.$

  \item In the model we assumed that $n_v$ was an increasing
    function. The results follow if $n_v$ is a decreasing function,
    except that $T^{\text{NE}}(v)$ now would be a non-decreasing
    function in $v.$ This suits our intuition as now the cost of
    waiting is more for customers with smaller $v.$
\end{enumerate}

\subsection{A Server with a Continuum of Service Grades}
\label{sec:continuum-grades}


In the preceding, the customer had to choose the time of arrival and
that time determined its total cost. If we could treat time as just a
parameter we could have different interpretations to the parameter and
hence apply it to different systems. The following is an example of
such a generalization which results in a relatively different server
model but yields an interesting NE arrival profile.

A population of heterogeneous customers, characterized by a type
$\mathcal{V} = [A, B] \subset \mathbb{R}^{+},$ arrive for
service. They have to buy a priority $w$ from the server which is
priced by the server. Without loss of generality we assume the class
of all priorities to be $\mathcal{W} = [0,1]$ and $P:\mathcal{W}
\rightarrow \mathbb{R}^{+}$ to be the pricing function such that
$P(w)$ denotes the price for priority $w.$ We assume $P$ to be
  any continuous, differentiable, monotonically increasing function
  with $P(0) = 0$ and $P(1) = P_{\text{max}}.$ Then the cost incurred
by a customer of type $v$ when she buys priority $w$ is given by
\begin{equation}
  \label{eq:cost_single_p}
  c_{v}(w) =  N F(w) + n_{v}P(w),
\end{equation}
where $n_v$ is an increasing function in $v$ and $F(w)$ is the
fraction of customers who choose a priority higher than $w.$ Note that
$n_{v}P(w)$ is the cost incurred by the type $v$ customer in buying
priority $w.$

Since, this is only a different interpretation of our original system
model, the results of Lemma~\ref{lem:ne_property_single_p} and
Theorem~\ref{thm:ne_single_p} hold in this scenario too. Thus, the
unique NE is given by
\begin{equation}
  \label{eq:local_2}
  w^{\text{NE}}(v) = P^{-1}\left( \int_{v}^{B}N\frac{ d\mathcal{G}(x)
  }{n_x}\right).
\end{equation}

Let us now compute the expected revenue earned by the service
provider, and see what price function $P$ will maximize it. For a
given price function $P(\cdot),$ the revenue of the service provider
is
\begin{displaymath}
  R(P) = \int_{A}^B P\left(w^{\text{NE}}(v)\right) d\mathcal{G}(v).
\end{displaymath}
Substituting~\eqref{eq:local_2}, we obtain
\begin{displaymath}
  R(P) = \int_{A}^B \left( \int_{v}^{B}N\frac{d\mathcal{G}(x)}{n_{x}} \right) d\mathcal{G}(v).
\end{displaymath}

Observe that $R(P)$ defined above is independent of $P(\cdot).$ Thus,
when the customers are strategic and $\mathcal{G}(v)$ distribution is
known to the customers, the revenue to the service provider is
invariant to the pricing function that is increasing in the priorities
and has range with range $[0,P_{\max}]$! This leads us to
argue there is only so much willingness in the market to pay, and that
any pricing function will fully extract it. We argue that this
  is because with a continuous pricing function, it is possible to
  achieve perfect discrimination which in turn leads to revenue
  maximisation for the service provider \cite{Courcoubetis03}.  In
  contrast, we will see in the next two sections that a finite number
  of service levels does not extract the maximum revenue. We argue
  that this is because of the `quantisation' effect which in turn does
  not admit perfect discrimination because the customers will have to
  choose the `nearest' grade of service. 


\section{Multiple Queues: Customers Choose Arrival Time and Priority}
\label{sec:multiple-queues}

Now we consider the system of the previous section but with $L$ FCFS
queues. Queue~$l+1$ has strict priority over all queues of priority
level $l$ or less, i.e., it is served before queue~$l$ for all $l = 0$
to $L-1.$ Customers who join queue~$l$ have to pay an admission price
of $P_{l}.$ Since queue~$l+1$ has priority over queue~$l$ the server
enforces that $P_{l+1} > P_{l}.$ Queue~$0,$ however, has no admission
price, i.e., $P_{0} = 0.$

Let $c_{v}(l,t)$ denote the cost of joining queue~$l$ at time $t$
before boarding begins. Then they are defined as follows:
\begin{equation}
  c_{v}(l,t) = m_{v}P_{l} + N\sum_{j=l+1}^{L-1}F_{j}(0) + NF_{l}(t) +
  n_v g(t),
\end{equation}
where $F_{l}(t),$ for $0 \leq l \leq L-1,$ denote the fraction of
customers that wait for $t$ or longer in queue~$l.$ Further, $m_{v}P$
is how much a customer $v$ values price $P$ in relation to his rank
in the queue.

Each customer has to determine which queue she wants to join and when
to arrive. First, consider a customer of type $v$ who wants to join
Queue $l.$ Clearly, her optimal arrival time $T_{l}(v)$ is
\begin{equation}
  \label{eq:opt_time_2q}
  T_{l}(v) = \argmin_{t \geq 0} c_{v}(l, t).
\end{equation}
For this choice of $T_{l}(v),$ the cost of joining Queue $l$ will be
$c_{l}(v) \triangleq c_{v}(l, T_{l}(v)).$ Thus the optimal randomized
choice of queue is obtained from
\begin{equation}
  \label{eq:opt_dec_2q}
  q(v) = \argmin_{q_{j}} \sum_{l \in \mathcal{L}} q_{l} c_{v}(l, T_{l}(v)),
\end{equation}
where $q(v) = (q_{0}(v), q_{1}(v), \ldots, q_{L-1}(v))$ and $q_{l}(v)$
denotes the probability that a customer of type $v$ would join Queue
$l.$ An equilibrium strategy is defined as follows.
\begin{definition}
  \label{def:ne_complex}
  $\left(T_{l}(v), q_{l}(v)\right)_{l=0}^{L-1}$ is a NE policy if
  \eqref{eq:opt_time_2q} and \eqref{eq:opt_dec_2q} are satisfied,
  where for all $l \in \mathcal{L},$
  \begin{align}
    \label{eq:Fl_2q}
    F_{l}(t) = \int_{v \in \mathcal{V}} \mathbbm{1}_{T_{l}(v) \geq t}
    \ q_{l}(v) \ d \mathcal{G}(v)
  \end{align}
  is the fraction of customers that join Queue $l$ and arrive at least
  $t$ units of time before $0.$
\end{definition}
Let $\left(T_{l}^{\text{NE}}(v),
q_{l}^{\text{NE}}(v)\right)_{l=0}^{L-1}$ denote an equilibrium
strategy. We first analyze the system with a single queue, i.e., $L =
1,$ and then proceed to the analysis when there are $L$ queues.

Lemma~\ref{lem:ne_property_single_p} is first extended to $L$ queues
to show that at NE, the optimal joining times at each queue,
$T_{l}^{\text{NE}}(v),$ is non increasing in $v.$
\begin{lemma}
  \label{lem:ne_Lq_prop}
  $T_{l}^{\text{NE}}(v)$ is a non increasing function in $v$ for each
  $l \in \mathcal{L}.$
\end{lemma}
\proof{See Appendix~\ref{app:pf:lem:ne_Lq_prop}. \hfill \qed}

Recall that $A$ and $B$ are, respectively, the minimum and the maximum
values of the support of $v.$ Thus an implication of this lemma is
that at NE, type $A$ customers will be the first to arrive while type
$B$ customers will be the last to arrive and this is independently of
the queue that they join. This leads us to conclude that\footnote{The
  proofs of~\eqref{eq:ss1} and~\eqref{eq:ss2} are trivial.}
\begin{equation}
  T_{l}^{\text{NE}}(B) = 0, \label{eq:ss1}
\end{equation}
and
\begin{equation}
  F_{l}\left( T_{l}^{\text{NE}}(A)\right) = 0, \label{eq:ss2}
\end{equation}
for every $l \in \mathcal{L}.$

We now see how $T_{l}^{\text{NE}}(v)$ would vary from the case when
there is only a single queue to that when there are two or more
queues. Intuitively, when there are two or more queues, customers
would want to come later than they would if there was only a single
queue. This is confirmed by the following lemma.
\begin{lemma}
  \label{lem:ne_Lp_prop_two}
  For all $l$, we have $T_{l}^{\text{NE}}(v) \leq g^{-1}\left(N \int_{v}^{B}
    \frac{d\mathcal{G}(x)}{n_{x}}\right).$
\end{lemma}
\proof{See Appendix~\ref{app:pf:lem:ne_Lp_prop_two}. \hfill \qed}

From Theorem~\ref{thm:ne_single_p}, $g^{-1}\left(N \int_{v}^{B}
  \frac{d\mathcal{G}(x)}{n_{x}}\right)$ is the arrival time at NE in a single
queue system. Using the previous results, we next show that there is a
unique NE and at this NE, the strategies are pure. Toward that proof
the following assumptions will be made on the system parameters.
\begin{description}
\item{\textbf{(A1)}} We assume that $m_{v}$ and $n_{v}$ are such that
  $n_{v}' \int_{v}^{B}\frac{d \mathcal{G}(x)}{n_{x}}$ does not grow faster than
  $-m_{v}'.$ Specifically,
  \begin{equation}
    y(v) \triangleq \frac{n_{v}'}{\left(-m_{v}'\right)}
    \int_{v}^{B} \frac{d \mathcal{G}(x)}{n_{x}},  \nonumber
  \end{equation}
  is a bounded function over $v \in \mathcal{V}.$ Also,
  $\int_{A}^{B}\frac{d\mathcal{G}(x)}{n_{x}} < \infty.$

\item{\textbf{(A2)}} We also make an assumption on the minimum
  difference between the prices to join `adjacent'
  queues. Specifically, for every $l$ and $l+1$ in $\mathcal{L},$
  \begin{equation}
    P_{l+1} - P_l > N \max \left\{ \sup_{v \in \mathcal{V}} y(v),
      \frac{2}{m_{A}}\right\}.
  \end{equation}

\item{\textbf{(A3)}} Finally, we assume that $m_B$ is small enough so
  that all queues are occupied, and also that $P_{l+1} - P_{l} <
  \frac{NF_{l}(0)}{m_{B}}.$
\end{description}
The following example illustrates that (A1) and (A2) are not very
restrictive. Let $v$ be uniformly distributed over $[A, B],$ $n_{v} =
v,$ and
\begin{equation}
  m_{v} = \frac{N}{\epsilon(B-A)}\left( B\log\left(\frac{B}{v}\right) -
    (B-v)\right), \label{eq:xyz}
\end{equation}
where $\epsilon > 0.$ In this case,
\begin{equation}
  y(v) = \frac{\epsilon}{N} \frac{\log\left(\frac{B}{v}\right)}{\left(
      \frac{B}{v} - 1 \right)} \leq \frac{\epsilon}{N}. \nonumber
\end{equation}
Also, note that, if $m_{v}$ satisfies (A2) then so does
$m_{v} + \delta.$
Since $y(v) \leq \frac{\epsilon}{N},$ it suffices to have
\begin{equation}
  P_{l+1} - P_{l} > \max \left\{ \epsilon, \frac{N}{m_{A}}\right\},
  \nonumber
\end{equation}
to satisfy (A2). Further, if $A = 0$ then $m_A = \infty,$ in which
case we only require $P_{l+1} - P_{l} > \epsilon.$ In general, this
would be the case if $m_A$ is sufficiently large.  If (A3) is violated
the empty queues will be those of higher priority. From an operational
point of view there is no reason to have a queue if no customer is
going to join it, especially a high priced queue.

\begin{theorem}
  \label{thm:ne_Lq_main_existence}
  The NE strategy is unique and is characterized by
  \begin{equation}
    \label{eq:b1}
    q_{l}^{\text{NE}}(v) = \mathbbm{1}_{v_{l} < v \leq v_{l+1}}.
  \end{equation}
  Here $A = v_{0} < v_{1} < v_{2} < \cdots < v_{L-1} < v_{L} = B$ are
  given by
  \begin{equation}
    \nonumber
    c_{l-1}(v_{l}) = c_{l}(v_{l}),
  \end{equation}
  for all $l = 1$ to $L-1,$ each of which has a unique solution.
\end{theorem}
\proof{The outline of the proof is as follows.
  \begin{enumerate}
  \item Lemma~\ref{lem:opt_join_cost_incdec} first shows that for type
    $A$ customers, the optimal queue joining cost, $c_{l}(A),$
    increases in $l,$ whereas for type $B$ customers this cost,
    $c_{l}(B),$ decreases in $l.$
  \item Lemma~\ref{lem:opt_cost_prop} shows that $\frac{d c_{0}(v)}{d
      v} > 0,$ while for $l \geq 1,$ $0 > \frac{d c_{l}(v)}{d v} >
    \frac{d c_{l+1}(v)}{d v}.$
  \item Using this, in the third part of
    Lemma~\ref{lem:opt_cost_prop}, we show that for every $l$ the cost
    functions $c_{l-1}(v)$ and $c_{l}(v)$ intersect at a unique point
    $v_l$ and these thresholds $\left\{ v_l\right\}_{l=1}^{L-1}$
    satisfy
    \begin{equation}
      A < v_{1} < v_{2} < \cdots < v_{L-1} < B. \nonumber
    \end{equation}
  \item Given the thresholds $\left\{ v_l\right\}_{l=1}^{L-1},$ we
    show that the optimal joining cost for Queue $l$ is greater than
    those of other queues for all $v \in (v_{l}, v_{l+1}],$ i.e.,
$c_{l}(v) \leq c_{j}(v)$ for all $v \in (v_{l}, v_{l+1}]$ and $j \in
  \mathcal{L}.$\footnote{It does not matter whether the customer of
    type $v_{l}$ joins either Queue~$l$ or Queue~$l+1.$ Set of all
    such customers form a negligible (measure $0$ w.r.t. $\mathcal{G}$) set in
    $\mathcal{V}.$}
  \end{enumerate}
  This proves that at NE we have~\eqref{eq:b1}. The details are in
  Appendix~\ref{app:pf:thm:ne_Lq_main_existence}.  \hfill \qed}

We now give an explicit characterization of the NE arrival times,
$T_{l}^{\text{NE}}(v),$ and the thresholds, $\left\{ v_{l} \right\}_{l
  = 0}^{L},$ in the following theorem.
\begin{theorem}
  \label{thm:ne_Lq_main}
  At NE, the arrival time to Queue $l$ is given by
  \begin{equation}
    \nonumber
    T_{l}^{\text{NE}}(v) = \left\{
      \begin{array}{ll} 0, &
        \!\!\!\text{for}~~v > v_{l+1} \\
        g^{-1}\left(N \int_{v}^{v_{l+1}} \frac{d\mathcal{G}(x)}{n_{x}} \right), &
        \!\!\!\text{for}~~v_{l} \leq v \leq v_{l+1} \\
        T_{l}^{\text{NE}}(v_{l}), & \!\!\!\text{for}~~v < v_{l}
      \end{array}
    \right.\!\!\!.
  \end{equation}
  Further, the thresholds $\left\{ v_{l} \right\}_{l = 0}^{L}$ can be
  computed by solving
  \begin{equation}
    \label{eq:th_def}
    \mathcal{G}(v_{l-1}) = \mathcal{G}(v_{l+1}) - \left(\frac{P_{l} - P_{l-1}}{N}\right)
    m_{v_{l}} - n_{v_{l}} \int_{v_{l}}^{v_{l+1}}\frac{d \mathcal{G}(x)}{n_{x}},
  \end{equation}
  for $1 \leq l \leq L-1.$
\end{theorem}
\proof{See Appendix~\ref{app:pf:thm:ne_Lq_main}. \hfill \qed}

We illustrate the preceding results with an example. Consider a
boarding game with $\mathcal{V} = [0, 20]$ and the customer population
distributed uniformly over $\mathcal{V}.$ The service provider
operates three queues with respective prices $P_0 = 0,$ $P_1 = 8.75,$
and $P_2 = 11.45.$ Let $N = 10,$ $n_{v} = v,$ and $m_v$ modified
from~\eqref{eq:xyz} as follows.
\begin{equation}
  m_{v} = \frac{N}{\epsilon(B-A)}\left( B\log\left(\frac{B}{v}\right)
    - (B-v)\right) + \frac{N\delta}{\epsilon}. \label{eq:zyx}
\end{equation}
For this example, we choose $\delta = 0.05.$

Note that, we do not need the value of $\epsilon$ to determine
  the NE policy, as it can be subsumed in the prices $P_l$. Whenever
  we plot the price $P_l$ and revenue, they are normalized by
  $\epsilon.$

\begin{figure}
  \centering
  \includegraphics[width=0.85\linewidth]{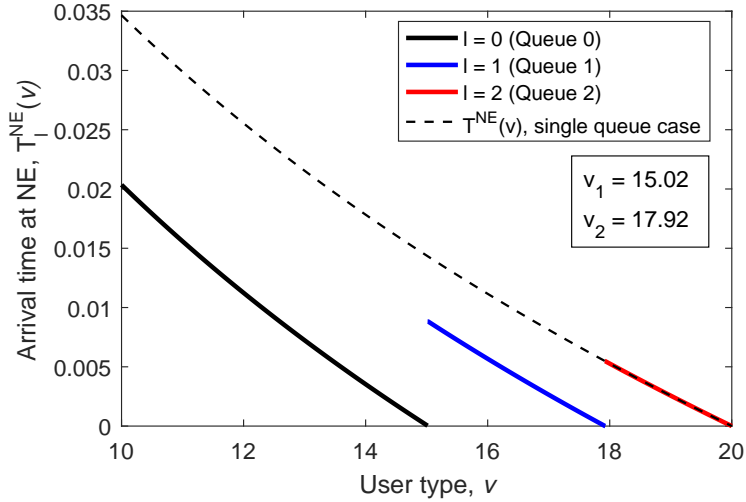}
  \caption{Comparison of NE arrival times at different queues for a
    system with three queues ($v \sim \mathcal{U}[0,20],$ $N = 10,$
    $P_1 = 8.75,$ $P_2 = 11.45$ and $\delta = 0.05$).}
  \label{fig:optimal_arrival_times_3q}
\end{figure}

Figure~\ref{fig:optimal_arrival_times_3q} plots the arrival times at
the three queues. Observe that $T_{l}^{\text{NE}}(v)$ is a decreasing
function of $v,$ which concurs with Lemma~\ref{lem:ne_Lq_prop}. In
equilibrium, we observe that customers of a lower priority start
arriving before the customers of higher priority. It is,
  however, not true that, in equilibrium, a customer which chooses a
  higher priority will arrive later than a customer that chooses a
  lower priority. We can observe this in Figure~1. For example,
  customer $v = 16$ chooses a higher priority than customer $v = 14$
  but arrives earlier than the customer $v = 14$.

Also plotted in Figure~\ref{fig:optimal_arrival_times_3q} is the
optimal arrival time as a function of the customer type when there is
only a single queue. We see that this arrival time is greater than the
arrival times at each queue for the three queue example. This concurs
with Lemma~\ref{lem:ne_Lp_prop_two}.  We only plot the arrival times
for $v \geq 10.$ For $v < 10$ the time to join the $0$-th queue
increases rapidly, however, it does remain upper bounded by the
optimal joining time if there was a single queue.

\begin{figure}
  \centering
  \includegraphics[width=0.82\linewidth]{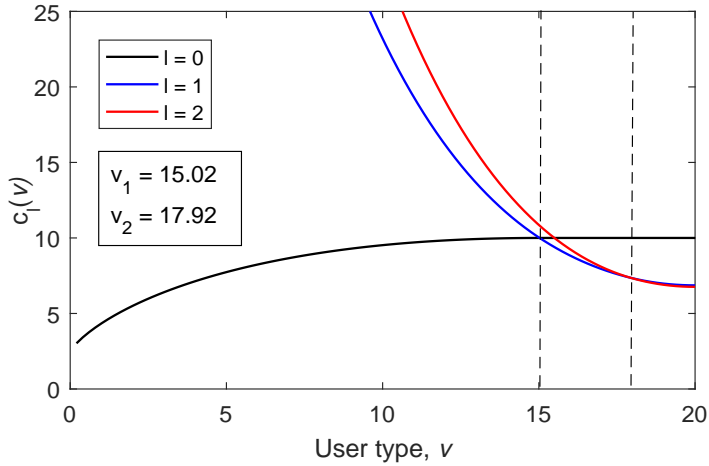}
  \caption{Comparison of optimal costs to join Queue $l$ and
    illustration of the thresholds ($v \sim \mathcal{U}[0,20],$ $N =
    10,$ $P_1 = 8.75,$ $P_2 = 11.45$ and $\delta = 0.05$).}
  \label{fig:optimal_joining_costs_3q}
\end{figure}
To illustrate the thresholds at NE, the optimal joining costs,
$c_{l}(v),$ for each queue is plotted as a function of $v$ in
Figure~\ref{fig:optimal_joining_costs_3q}. A customer will join
Queue~$l$ if it offers it the least optimal joining cost. Thus, the
crossing points between the optimal joining costs determine the
thresholds; this is observed clearly in
Figure~\ref{fig:optimal_joining_costs_3q}.

\subsection{Discussion}
\label{sec:multiple-discuss}

We now explore how the NE strategy depends on the system parameters,
specifically on the prices. We consider a two-queue boarding game with
customer types uniformly distributed over $[5, 15].$ $n_{v} = v$ and
$m_{v}$ is given by ~\eqref{eq:zyx}.

\begin{figure}
  \centering
  \includegraphics[width=0.90\linewidth]{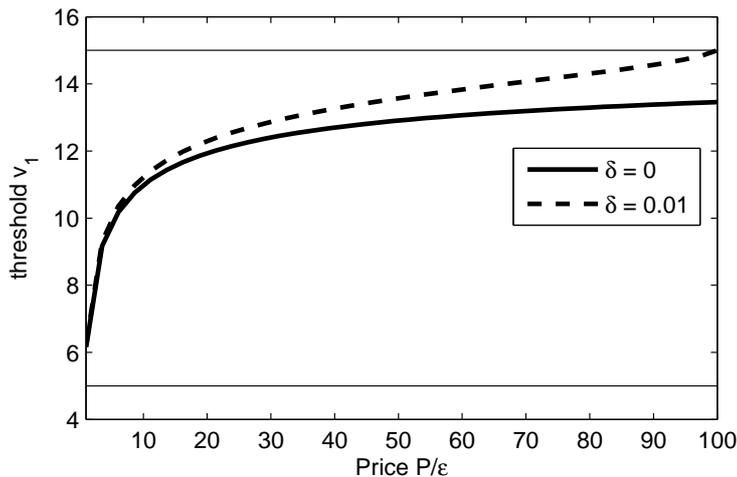}
  \caption{Threshold $v_{1}$ as a function of price $P$ for the
    two-queue boarding game ($v \sim \mathcal{U}[5, 15]$).}
  \label{fig:th_plot_1}
\end{figure}

Figure~\ref{fig:th_plot_1} plots the threshold $v_{1}$ as a function
of the price, $P,$ of the priority queue. Observe that $v_1$ increases
with $P.$ This is intuitive because the incentive to join the high
priority queue is diminished as the price increases. In fact, for
$\delta = 0.01,$ we find that $v_1 = B$ when $P = 100\epsilon.$
However, this is not true when $\delta = 0,$ in which case $v_1$ keeps
increasing to $B,$ but is never equal to it. Investigating this
further, since $NF_{1}(0)$ is the number of customers joining Queue
$1,$ revenue earned by the server would be $R(P) \triangleq
NF_{1}(0)P.$ This is not a linear function of $P$ as $F_{1}(0) =
\frac{B - v_{1}}{B - A}$ and depends on $P.$

\begin{figure}
  \centering
  \includegraphics[width=0.85\linewidth]{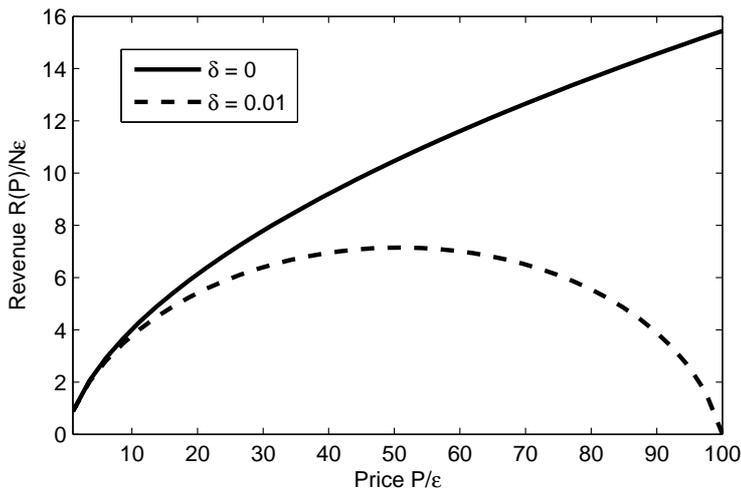}
  \caption{Revenue earned by the service provider, $R(P),$ as a
    function of the price $P$ for the two-queue boarding game ($v \sim
    \mathcal{U}[5, 15]$).}
  \label{fig:rev_ex4}
\end{figure}

In Figure~\ref{fig:rev_ex4}, we plot the revenue (normalized by
$N\epsilon$) as a function of the price $P$ (normalized by
$\epsilon$). For $\delta = 0,$ we observe that the revenue keeps
increasing as $P$ is increased, while for $\delta = 0.01$ there is an
optimal choice of $P$ at which the revenue attains its maximum. Note
that $\delta = 0$ (resp. $\delta > 0,$) corresponds to the case when
$m_B = 0$ (resp. $m_B > 0$). Hence, the intuition is that when $m_{B}
= 0,$ there are always customers (close to customers of type $B$) who
do not value money much. Therefore, the service provider can always
increase her revenue by setting a higher price for the priority
queue. On the contrary, when $m_{B} > 0,$ there exists a price $P,$
large enough, that customers of type $B$ cannot afford. The latter
decide to join Queue~$0.$

In the previous discussions, we have considered $m_{v}$ that decreases
more rapidly for smaller $v$ than it does for larger $v.$ Now consider
a system where $m_v$ decreases linearly, i.e.,
\begin{equation}
  m_{v} = \frac{N}{\epsilon}\left(\frac{B-v}{B-A}\right)\log(B/A)
  + \frac{N\delta}{\epsilon}.
\end{equation}
We now investigate the two-queue boarding game with $m_{v}$ as above,
$n_{v} = v,$ and $g(t) = t,$ and customer types uniformly distributed
over $[A, B].$ For this system $y(v)$ is
\begin{align}
  y(v) \ = \ \frac{1}{-m_{v}'} \int_{v}^{B}\frac{1}{B-A}\frac{dx}{x}
  \ =
  \ \frac{\epsilon}{N}\frac{\log\left(\frac{B}{v}\right)}
         {\log\left(\frac{B}{A}\right)}. \nonumber
\end{align}
For $y(v)$ to be bounded we require $A > 0,$ in which case $y(v) \leq
\frac{\epsilon}{N}.$ This upper bound is achieved when $v = A.$ As
observed previously, if $m_A$ is sufficiently large then for~(A2) it
is sufficient if $P_{l+1} - P_{l} > \epsilon.$ Further, for
$\int_{v}^{B}\frac{d \mathcal{G}(x)}{n_x}$ to be bounded, we require $B <
\infty.$
The thresholds and revenue behave as in the previous example

\section{Revenue Maximization}
\label{sec:revenue}

In the previous section, the admission price to each of the $L$ queues
was assumed given and we provided a complete characterization of the
NE strategy. In this section, we investigate how the service provider
can maximize its revenue by appropriately setting the prices $P_l.$
Our motivational setting is the airplane boarding system where
customers have already purchased their tickets. Hence it is reasonable
to assume that there is one queue, the lowest priority queue, where
there is no premium, i.e., $P_0=0.$ Thus although the customers cannot
balk, the existence of the a `free queue' ensures that the revenue
is not an increasing function of the premiums in the higher
priority queues and there will be an optimum value for these
prices. Further, increased service differentiation allows for
increased revenues. In this section we first determine the optimum
prices for a given $L$ and then numerically analyze the effect of
increasing $L.$

For the $L$ queue system, $F_{l}(0)$ is the fraction of customers that
joined Queue $l$ and, hence, $N F_{l}(0)$ is the total number of
customers that join Queue~$l.$ The revenue earned by the service
provider is given by
\begin{equation}
  R(P_{1}, P_{2}, \ldots, P_{L-1}) = \sum_{l=1}^{L-1} N F_{l}(0) P_{l}.
\end{equation}
We know that, at NE, $F_{l}(0) = \mathcal{G}(v_{l+1}) -
\mathcal{G}(v_{l}).$ Also,~\eqref{eq:th_def} gives a bijective relation between
$P_l$ and $v_{l}.$ Using these, we can express revenue as follows:
\begin{equation}
  \frac{1}{N}R(P_{1}, P_{2}, \ldots, P_{L-1}) = \sum_{j=1}^{L-1}
  u\left(v_{L-j}, v_{L-j+1}\right),
\end{equation}
where
\begin{multline}
  u\left(v_{l}, v_{l+1}\right) = \left[ \frac{\mathcal{G}(v_{l+1}) -
      v_{l}\int_{v_{l}}^{v_{l+1}}\frac{d\mathcal{G}(x)}{x} }{
      \frac{1}{N}m_{v_{l}} }\right] \left( 1 - \mathcal{G}(v_{l}) \right)
  \nonumber \\
  - \left[\frac{\mathcal{G}(v_{l})}{\frac{1}{N}m_{v_{l+1}}}\right] \left( 1 -
  \mathcal{G}(v_{l+1}) \right).
\end{multline}
This is derived in Appendix~\ref{app:der:rev_fun}.

Thus, maximizing $\frac{1}{N}R(P_{1}, P_{2}, \ldots, P_{L-1})$ can be
solved as a finite horizon dynamic program (FHDP). Notice that the
first term in the expansion depends only on $v_{L-1}$ (action at time
$1$ in the FHDP), the second term depends only on $v_{L-2}$ and
$v_{L-1}$ (action and state of the system at time $t = 2$), the third
term depends only on $v_{L-3}$ and $v_{L-2}$ (action and state of the
system at time $t = 3$), and so on. We assume that the FHDP has a
solution and verify this numerically.

\begin{table}
  \centering
  \caption{Revenue maximizing prices, and the corresponding
    thresholds, for two, three, and four queues ($v \sim
    \mathcal{U}[0, 150]$).}
  \begin{tabular}{| l || l | l | l | l | l | l |}
      \hline
    L & $v_1$ & $v_2$ & $v_3$ & $P_{1}$ & $P_{2}$ & $P_{3}$ \\ \hline \hline
    2 & $135.28$ & - & - & $76.73$ & - & - \\ \hline
    3 & $134.35$ & $143.46$ & - & $71.68$ & $79.75$ & - \\ \hline
    4 & $134$ & $139.22$ & $142.52$ & $69.92$ & $73.65$ & $76.58$ \\
    \hline
    \end{tabular}
\label{tbl:th_price}
\end{table}

We now numerically evaluate the revenue maximizing prices by solving
the FHDP. As before, $\mathcal{V} = [A, B]$ and type is uniformly
distributed. There are $L$ queues, $N = 10,$ $n_{v} = v,$ and $m_{v}$
is as in~\eqref{eq:zyx} with $\delta = 0.05.$

Table~\ref{tbl:th_price} lists the revenue maximizing prices and
thresholds when $A = 0$ and $B = 150$ for $L = 2,$ $3,$ and $4.$ We
observe that as the number of queues is increased, for maximum
revenue, $P_l$ has to be decreased at each Queue $l.$ This has been
consistently observed for other system parameters.

\begin{table}
  \centering
  \caption{Maximum revenue as a function of number of queues and type distribution}
  \begin{tabular}{|l||l|l|l|}
    \hline
    Population &\multicolumn{3}{c|}{Revenue}\\ \cline{2-4}
    distribution & $L=2$ & $L=3$ & $L=4$ \\
    \hline\hline
    $\mathcal{U}[0,20]$ & $2.26$ & $2.46$ & $2.50$ \\ \hline
    $\mathcal{U}[0,150]$ & $7.53$ & $7.83$ & $7.87$ \\ \hline
    $\mathcal{U}[20,150]$ & $7.41$ & $7.65$ & $7.68$ \\ \hline
  \end{tabular}
  \label{tbl:rev}
\end{table}

In Table~\ref{tbl:rev}, we compare the maximum revenue for three
different type distributions when $L = 2,$ $3,$ and $4.$ We observe
that the revenue increases as more queues are added, regardless of the
underlying type distribution. Most importantly, the increase in this
revenue from three queues to four queues is as small as $1.6\%,$
$0.5\%,$ and $0.4\%$ for the type distributions $\mathcal{U}[0,20],$
$\mathcal{U}[0,150],$ and $\mathcal{U}[20,150],$
respectively. However, for the same type distributions the increase in
maximum revenue is $8.9\%,$ $4\%,$ and $3.2\%,$ respectively. This
tells us that having three, may be even just two, queues
shall get us very nearly the maximum revenue.

We also remark here that this observation of a small number of
  classes yielding nearly the maximum revenue is also seen in other
  similar settings that serve a heterogenous population. For example,
  in \cite{Shakkottai08}, the authors consider dividing link capacity
  into multiple classes of service, each with its own price, an
  instance of Paris Metro pricing. Utility maximising users choose the
  class of service and it is shown that for a large range of utility
  functions, the loss of revenue is small even with a small number of
  service classes.

Another observation from the results of Table~\ref{tbl:rev} is that
the revenue is larger when the type distribution is
$\mathcal{U}[0,150].$ Numerically, it is consistently observed that if
$\tilde{\mathcal{V}} \subset \mathcal{V},$ then the boarding game with
types in $\mathcal{V}$ yields greater maximum revenue than with
$\tilde{\mathcal{V}};$ a larger diversity of types provides a larger
revenue. This property is known in the pricing literature,
  e.g., \cite{Courcoubetis03}. Larger diversity provides for more
  customers with higher valuations for quality of service (QoS).  Such
  a user population allows the service provider to extract more
  consumer surplus by tuning service quality and price to each type to
  maximise profit.

\section{Concluding Discussion}
\label{sec:conclusion}

The system that we have introduced in this paper can be seen to belong
to a class of queueing systems where the arrival times are endogenously
determined by the customers; unlike the exogenously determined arrival
times in traditional queueing systems. Clearly, there are several potential
applications for systems with endogenous arrivals in modeling waiting
systems at airports, bus and train terminals, concert halls, etc.  A better
understanding of these systems can help in a more informed sizing
of waiting facilities.

Several extensions are possible. An extension of immediate
interest is to develop a learning algorithm to obtain the revenue
maximizing prices. Specifically, since such a game would be played
out repeatedly, the outcome of each instance may be used to adapt the
prices to maximize the revenue. A second extension would be to allow
customers to balk if the cost is higher than the value of obtaining service.


\appendix

\section{Proof of Lemma~\ref{lem:ne_property_single_p}}
\label{app:pf:lem:ne_property_single_p}

The proof uses the optimality of $T^{\text{NE}}(v)$ and
$T^{\text{NE}}(v+h)$ for $c_{v}(\cdot)$ and $c_{v+h}(\cdot),$
respectively, and the structure of the cost function.

\begin{align*}
  c_{v}(T^{\text{NE}}(v)) &\leq c_{v}\left(T^{\text{NE}}(v+h)\right)\\
  &= N F\left(T^{\text{NE}}(v+h)\right) + n_{v}g\left(T^{\text{NE}}(v+h)\right) \\
  &= c_{v+h}\left(T^{\text{NE}}(v+h)\right) - n_{v+h}g\left(T^{\text{NE}}(v+h)\right) + n_{v}g\left(T^{\text{NE}}(v+h)\right).
\end{align*}
Similarly, we can get
\begin{equation*}
  c_{v+h}(T^{\text{NE}}(v+h)) \leq c_{v}\left(T^{\text{NE}}(v)\right) -
  n_{v}g\left(T^{\text{NE}}(v)\right) + n_{v+h}g\left(T^{\text{NE}}(v)\right).
\end{equation*}
Adding the two we obtain
\begin{displaymath}
  \left( n_{v+h} - n_{v} \right)\left( g\left(T^{\text{NE}}(v+h)\right)
    - g\left(T^{\text{NE}}(v)\right) \right) \leq 0.
\end{displaymath}
Note that $n_{v+h} > n_{v},$ as $n_{v}$ is a strictly increasing function. Therefore,
\begin{displaymath}
  g\left(T^{\text{NE}}(v+h)\right) \leq  g\left(T^{\text{NE}}(v)\right).
\end{displaymath}
This implies that $T^{\text{NE}}(v+h) \leq T^{\text{NE}}(v),$ as
$g(\cdot)$ is a strictly increasing function.

\section{Proof of Theorem~\ref{thm:ne_single_p}}
\label{app:pf:thm:ne_single_p}

First note that $F(T^{\text{NE}}(v)) = \mathcal{G}(v).$ This is derived
using~\eqref{eq:def_ft} as follows.
\begin{equation}
  F(T^{\text{NE}}(v)) = \int \mathbbm{1}_{T^{\text{NE}}(x) >
    T^{\text{NE}}(v)} d\mathcal{G}(x) = \int \mathbbm{1}_{x < v} d\mathcal{G}(x) = \mathcal{G}(v), \label{eq:aa1}
\end{equation}
because $\mathcal{G}$ is a non-atomic distribution, and the second equality follows from Lemma~\ref{lem:ne_property_single_p}.

Now, since $T^{\text{NE}}(v)$ is a NE, by first order condition
on~\eqref{eq:cost_single_p}, we have
\begin{align}
  0 &= \left. \frac{d c_{v}(t)}{d t} \right|_{t = T^{\text{NE}}(v)}
  \nonumber \\
  &= \left. N \frac{d F(t)}{d
      t}\right|_{t = T^{\text{NE}}(v)} + n_{v}g'\left(T^{\text{NE}}(v)\right), \label{eq:a1}
\end{align}
where $g'(x)$ denotes derivative $g$ with respect to $x.$ Differentiating~\eqref{eq:aa1} with
respect to $v$ we get
\begin{equation}
  \frac{d \mathcal{G}(v)}{d v} = \frac{d F(T^{\text{NE}}(v))}{d v} = \frac{d T^{\text{NE}}(v)}{d v} \times \left. \frac{d F(t)}{d t}\right|_{t = T^{\text{NE}}(v)}. \label{eq:a2}
\end{equation}
Substituting~\eqref{eq:a2} in~\eqref{eq:a1}, we obtain
\begin{equation}
  - \frac{N}{n_{v}} \frac{d
    \mathcal{G}(v)}{d v} = g'\left(T^{\text{NE}}(v)\right) \frac{d
    T^{\text{NE}}(v)}{d v} = \frac{d g\left( T^{NE}(v)\right)}{d v}. \nonumber
\end{equation}
Integrating both sides with respect to $v,$ we get
\begin{align}
  \int_{v}^{B} d g\left( T^{NE}(x)\right) = - \int_{v}^{B} \frac{N}{n_x} d\mathcal{G}(x).
\end{align}
This implies
\begin{equation}
\label{eq:aa2}
g\left( T^{NE}(B)\right) - g\left(T^{NE}(v)\right) = - \int_{v}^{B} \frac{N}{n_x} d\mathcal{G}(x).
\end{equation}
Note that $T^{NE}(v)$ is a non-increasing function. Thus, users of type $B$ arrive last in the queue. If $T^{NE}(B) > 0$ then the cost for user of type $B$ can be reduced by decreasing $T^{NE}(B)$ to $0$, which contradicts the definition of $T^{NE}(v)$. Therefore, $T^{NE}(B) = 0$. This implies that $g\left( T^{NE}(B)\right) = 0$ due to assumptions on $g$. \eqref{eq:aa2} then implies the result.

\section{Proof of Lemma~\ref{lem:ne_Lq_prop}} 
\label{app:pf:lem:ne_Lq_prop}
For notational simplicity, we shall denote $D(l,t)$ to mean
\begin{equation}
NF_{l}(t) + \sum_{j=l+1}^{L-1} NF_{j}(0),
\end{equation}
in this section. Thus, the cost of a customer $v$ is given by
\begin{equation}
c_{v}(l,t) = D(l,t) + m_{v}P_l + n_{v}g(t).
\end{equation}
Comparing the costs of customer $v$ and $v+h,$ we have
\begin{equation}
\label{eq:th_pf_1}
c_{v+h}(l, t) = c_{v}(l, t) - m_{v}P_l - n_{v}g(t) + m_{v+h}P_{l+1} + n_{v+h}g(t).
\end{equation}
Substituting $t = T_{l}^{\text{NE}}(v)$ in~\eqref{eq:th_pf_1}, we obtain
\begin{align}
  c_{v+h}\left(l, T_{l}^{\text{NE}}(v)\right) &= c_{v}\left(l, T_{l}^{\text{NE}}(v)\right) - m_{v}P_{l} - n_{v}g\left(T_{l}^{\text{NE}}(v)\right) + m_{v+h}P_{l+1} \nonumber \\
  &~~~~~~~~~~~~~~~~~~~~~~~~~~~~~~~~~~~~~~+ n_{v+h}g\left(T_{l}^{\text{NE}}(v)\right) \nonumber \\
  &\geq c_{v+h}\left(l, T_{l}^{\text{NE}}(v+h)\right). \label{eq:th_pf_2}
\end{align}
The last inequality follows because $T_{l}^{\text{NE}}(v+h)$ minimizes
$c_{v+h}(l,\cdot).$
Similarly, substituting $t = T_{l}^{\text{NE}}(v+h)$
in~\eqref{eq:th_pf_1} we get
\begin{multline}
  c_{v+h}\left(l, T_{l}^{\text{NE}}(v+h)\right) = c_{v}\left(l, T_{l}^{\text{NE}}(v+h)\right) - m_{v}P_{l}
  - n_{v}g\left(T_{l}^{\text{NE}}(v+h)\right) + m_{v+h}P_{l+1} \\
  + n_{v+h}g\left(T_{l}^{\text{NE}}(v+h)\right).
\end{multline}
Since, $c_{v}\left(l, T_{l}^{\text{NE}}(v+h)\right) \geq c_{v}\left(l, T_{l}^{\text{NE}}(v)\right),$ we obtain
\begin{multline}
  c_{v+h}\left(l, T_{l}^{\text{NE}}(v+h)\right) \geq c_{v}\left(l, T_{l}^{\text{NE}}(v)\right) - m_{v}P_{l}
  - n_{v}g\left(T_{l}^{\text{NE}}(v+h)\right) + m_{v+h}P_{l+1} \\
  + n_{v+h}g\left(T_{l}^{\text{NE}}(v+h)\right), \label{eq:th_pf_3}
\end{multline}
Adding the two inequalities, namely,~\eqref{eq:th_pf_2}
and~\eqref{eq:th_pf_3}, we get
\begin{equation}
  \left(n_{v+h} - n_{v}\right)g\left(T_{l}^{\text{NE}}(v)\right) \geq \left(n_{v+h} - n_{v}\right)g\left(T_{l}^{\text{NE}}(v+h)\right). \label{eq:th_pf_4}
\end{equation}
Since, $n_v$ is an increasing function of $v,$~\eqref{eq:th_pf_4}
reduces to $g\left(T_{l}^{\text{NE}}(v)\right) \geq
g\left(T_{l}^{\text{NE}}(v+h)\right),$ which is nothing but
\begin{equation}
T_{l}^{\text{NE}}(v) \geq T_{l}^{\text{NE}}(v+h), \nonumber
\end{equation}
as $g(\cdot)$ is an increasing function. Thus, $T_{l}^{\text{NE}}(v)$ is a decreasing
function in $v.$

\section{Proof of Lemma~\ref{lem:ne_Lp_prop_two}}
\label{app:pf:lem:ne_Lp_prop_two}
By Definition~\ref{def:ne_complex},
\begin{equation}
  F_{l}\left( T_{l}^{\text{NE}}(v)\right) = \int \mathbbm{1}_{T_{l}^{\text{NE}}(x) \geq T_{l}^{\text{NE}}(v)} q_{l}(v) d\mathcal{G}(v) = \int \mathbbm{1}_{x \leq v} q_{l}(v) d\mathcal{G}(v),
\end{equation}
where the last equality follows due to Lemma~\ref{lem:ne_Lq_prop}. Now define
\begin{equation}
G_{l}(v) \triangleq \int_{A}^{v} q_{l}(v) d\mathcal{G}(v). \label{eq:mid_4}
\end{equation}
Then, we have $F_{l}\left( T_{l}^{\text{NE}}(v)\right) = G_{l}(v);$
also note that $G_{l}(v) \leq \mathcal{G}(v)$ for every $v.$ Differentiating
this w.r.t. $v,$ we obtain
\begin{equation}
\left. \frac{d F_{l}(t)}{d t} \right|_{t = T_{l}^{\text{NE}}(v)}\!\!\! \frac{d T_{l}^{\text{NE}}(v)}{dv} = \frac{d G_{l}(v)}{d v}. \label{eq:mid_3}
\end{equation}
Also, the first order derivative condition for the optimality of $T_{l}^{\text{NE}}(v),$ namely, $\frac{d c_{v}(l,t)}{d t} = 0,$ gives
\begin{align}
0 = \left[ N \frac{d F_{l}(t)}{d t} + n_{v}\frac{d g(t)}{d t} \right]_{t = T_{l}^{\text{NE}}(v)}. \label{eq:mid_5}
\end{align}
Substituting~\eqref{eq:mid_3} in~\eqref{eq:mid_5}, we get
\begin{equation}
\frac{d g\left(T_{l}^{\text{NE}}(v)\right)}{d v} = - \frac{N}{n_{v}} \frac{d G_{l}(v)}{d v}.
\end{equation}
This can be simplified to
\begin{align}
T_{l}^{\text{NE}}(v) = g^{-1}\left( N \int_{v}^{B} \frac{d G_{l}(x)}{n_x}\right) \leq g^{-1}\left( N \int_{v}^{B} \frac{d \mathcal{G}(x)}{n_x}\right),
\end{align}
where, while the second inequality follows from $G_{l}(v) \leq \mathcal{G}(v),$ the first equality uses~\eqref{eq:ss1} and the same arguments that are used in Appendix~\ref{app:pf:thm:ne_single_p} to arrive at the explicit expression of $w^{\text{NE}}(v).$

\section{Proof of Theorem~\ref{thm:ne_Lq_main_existence}}
\label{app:pf:thm:ne_Lq_main_existence}
For the ease of presentation, we denote
\begin{equation}
D(l,t) \triangleq N F_{l}(t) + N\sum_{j=l+1}^{L-1}F_{j}(0).
\end{equation}
In the following lemma, we prove that $c_{l}(A)$ increases, while $c_{l}(B)$ decreases in $l.$
\begin{lemma}
\label{lem:opt_join_cost_incdec}
$c_{l}(A)$ strictly increases and $c_{l}(B)$ strictly decreases in $l.$
\end{lemma}
\proof{The optimal cost to join Queue $l$ is given by
\begin{equation}
c_{l}(v) = D\left(l,T_{l}^{\text{NE}}(v)\right) + m_{v}P_{l} + n_{v}g\left(T_{l}^{\text{NE}}(v)\right). \nonumber
\end{equation}
At $v = A,$
\begin{align}
c_{l}(A) &= D\left(l,T_{l}^{\text{NE}}(A)\right) + m_{A}P_l + n_{A}g\left(T_{l}^{\text{NE}}(A)\right) \nonumber \\
&= N\sum_{j = l+1}^{L-1} F_{j}(0) + m_{A}P_l + n_{A}g\left(T_{l}^{\text{NE}}(A)\right), \nonumber
\end{align}
where the last equality follows from~\eqref{eq:ss2}. We therefore have
\begin{align}
c_{l+1}(A) - c_{l}(A) &= -N F_{l+1}(0) + m_{A}\left( P_{l+1} - P_{l}\right) - n_{A} \left( g\left(T_{l}^{\text{NE}}(A)\right) - g\left(T_{l+1}^{\text{NE}}(A)\right)\right) \nonumber \\
&\geq - N + m_{A}\left( P_{l+1} - P_{l}\right) - n_{A}g\left(T_{l}^{\text{NE}}(A)\right) \nonumber \\
&\geq - N + m_{A}\left( P_{l+1} - P_{l}\right) - N n_{A} \int_{A}^{B} \frac{d\mathcal{G}(x)}{n_{x}}, \label{eq:arbit}
\end{align}
where the last inequality follows from Lemma~\ref{lem:ne_Lp_prop_two} and the fact that $g(\cdot)$ is an increasing function. Using the fact that $n_{x}$ is a decreasing function and $\int_{A}^{B}\frac{d\mathcal{G}(x)}{n_{x}} < \infty$ from~(A1), we have
\begin{equation}
\int_{A}^{B}\frac{n_{A}}{n_{x}}d\mathcal{G}(x) \leq \int_{A}^{B}d\mathcal{G}(x) = 1. \nonumber
\end{equation}
Using this in~\eqref{eq:arbit} gives
\begin{equation}
c_{l+1}(A) - c_{l}(A) \geq - 2N + m_{A}\left( P_{l+1} - P_{l}\right) > 0, \nonumber
\end{equation}
where the last inequality follows from (A2).

At $v = B,$ we have
\begin{equation}
c_{l}(B) = D\left(l,T_{l}^{\text{NE}}(B)\right) + m_{B}P_l + n_{B}g\left(T_{l}^{\text{NE}}(B)\right) = N\sum_{j = l}^{L-1} F_{j}(0) + m_{A}P_l, \nonumber
\end{equation}
where the last equality follows from~\eqref{eq:ss1} and the fact that $g(0) = 0.$ Now,
\begin{equation}
c_{l+1}(B) - c_{l}(B) = -N F_{l}(0) + m_{B}\left( P_{l+1} - P_l\right) < 0,
\end{equation}
from~(A3). \hfill \qed}

We now prove some properties of the optimal costs $c_{l}(v)$ which help us in arriving at the thresholds.
\begin{lemma}
\label{lem:opt_cost_prop}
For the optimal costs, $c_{l}(v)$s, the following is true
\begin{enumerate}
  \item $\frac{d c_{0}(v)}{d v} > 0.$
  \item For all $l \in \{1, 2, \ldots L-2\},$ we have $0 > \frac{d c_{l}(v)}{d v} > \frac{d c_{l+1}(v)}{d v}.$
  \item There exists a unique point $v_{l},$ for every $l,$ at which the two costs, namely, $c_{l-1}(v)$ and $c_{l}(v)$ intersect. Also, $A < v_{1} < \cdots < v_{L-1} < B.$
\end{enumerate}
\end{lemma}

\proof{ Taking the derivative of $c_{l}(v)$ w.r.t. $v$ we get
\begin{equation}
\frac{d c_{l}(v)}{d v} = m_{v}' P_l + n_{v}' g\left( T_{l}^{\text{NE}}(v) \right) + \frac{d T_{l}^{\text{NE}}(v)}{dv} \times\left[ \frac{\partial D(l,t)}{\partial t} + n_{v} \frac{d g(t)}{d t}\right]_{t = T_{l}^{\text{NE}}(v)}. \label{eq:arbit2}
\end{equation}
Using the first order optimality condition for $T_{l}^{\text{NE}}(v),$ which is
\begin{equation}
\left[ \frac{\partial D(l,t)}{\partial t} + n_{v} \frac{d g(t)}{d t}\right]_{t = T_{l}^{\text{NE}}(v)} = 0,
\end{equation}
in~\eqref{eq:arbit2} we get
\begin{equation}
\frac{d c_{l}(v)}{d v} = m_{v}' P_l + n_{v}' g\left( T_{l}^{\text{NE}}(v) \right). \label{eq:mid_1}
\end{equation}
For $l = 0,$ since $P_0 = 0,$ we have
\begin{equation}
\frac{d c_{0}(v)}{d v} = n_{v}' g\left( T_{l}^{\text{NE}}(v) \right) > 0,
\end{equation}
which follows from the fact that $n_{v}$ is an increasing function in $v$ and $g(\cdot)$ always positive.

Now, for $l \geq 1,$ using Lemma~\ref{lem:ne_Lp_prop_two} in~\eqref{eq:mid_1} and the fact that $g(\cdot)$ is an increasing function, we get
\begin{equation}
\frac{d c_{l}(v)}{d v} < m_{v}' P_l + n_{v}'N\int_{v}^{B}\frac{d\mathcal{G}(x)}{n_{x}} = \left(-m_{v}'\right)\left( Ny(v) -  P_l \right) < 0,
\end{equation}
where the last inequality follows from the fact that $-m_{v}' > 0$ and (A2): since $P_j - P_{j-1} > N y(v)$ implies
\begin{equation}
P_l > N l y(v) \geq N y(v). \nonumber
\end{equation}

Analyzing the difference between $\frac{d c_{l}(v)}{d v}$ and $\frac{d c_{l+1}(v)}{d v},$ we obtain
\begin{align}
\frac{d c_{l}(v)}{d v} - \frac{d c_{l+1}(v)}{d v} &= -m_{v}'\left( P_{l+1} - P_l \right) + n_{v}' g\left(T_{l}^{\text{NE}}(v)\right) \nonumber \\
&~~~~~~~~~~~- n_{v}' g\left(T_{l+1}^{\text{NE}}(v)\right)  \nonumber \\
&> -m_{v}'\left( P_{l+1} - P_{l}\right) - n_{v}'g\left(T_{l+1}^{\text{NE}}(v)\right) \nonumber \\
&> -m_{v}'\left( P_{l+1} - P_{l}\right) - n_{v}'N\int_{v}^{B}\frac{d\mathcal{G}(x)}{n_{x}} \nonumber \\
&= -m_{v}\left[ \left( P_{l+1} - P_{l}\right) - Ny(v) \right] > 0,
\end{align}
where the last inequality follows from (A2), while the third in the
third step we use Lemma~\ref{lem:ne_Lp_prop_two} and the property that
$g(\cdot)$ is increasing. This proves part~$1$ and~$2$ of the Lemma.

For part~$3,$ take $l-1,$ $l,$ and $l+1$ in $\mathcal{L}.$ First note
that $c_{l-1}(A) < c_{l}(A)$ and $c_{l-1}(B) > c_{l}(B)$ from
Lemma~\ref{lem:opt_join_cost_incdec}. Since, $c_{l}(v)$ is continuous
and $\frac{d c_{l}(v)}{d v} > \frac{d c_{l+1}(v)}{d v},$ there exists
exactly one $v_{l}$ at which the two functions, namely, $c_{l-1}(v)$
and $c_{l}(v),$ meet. Further, note that $A < v_{1}$ because $c_{0}(A)
< c_{1}(A)$ and, similarly, $v_{L-1} < B$ because $c_{L-2}(B) >
c_{L-1}(B)$ by Lemma~\ref{lem:opt_join_cost_incdec}.

It only remains to show that $v_{l} < v_{l+1}.$ Using part~$1$ and~$2$
of the Lemma it is clear that for $v < v_{l}$ we have $c_{l-1}(v) <
c_{l}(v)$ and for $v > v_{l}$ we have $c_{l}(v) < c_{l-1}(v).$ Now, if
$v_{l+1} \leq v_{l}$ then for all $v \in \left(v_{l+1} B\right),$ and
hence for all $v \in \left[v_{l}, B\right),$ $c_{l+1}(v) <
c_{l}(v).$ Thus, over the interval $\left[A, v_{l}\right)$ we shall
have $c_{l-1}(v) < c_{l}(v)$ and over the interval $\left[v_{l},
  B\right)$ we shall have $c_{l+1}(v) < c_{l}(v).$ This implies that
no customer will join Queue $l$ contradicting (A3). \hfill \qed}

We now prove that joining Queue $l$ is the best strategy for all $v
\in \left(v_{l}, v_{l+1} \right].$ For this, it needs to be verified
that $c_{l}(v) \leq c_{j}(v)$ for $v \in \left(v_{l}, v_{l+1} \right]$
for all $j \in \{0, 1, \ldots, L-1\}.$ The following lemma provides
some sufficient conditions for it. We later show that the optimal cost
functions $c_{l}(v)$ satisfy these sufficient conditions.
\begin{lemma}
\label{lem:cond_c}
If
\begin{equation}
\label{eq:b2}
c_{l}(v) \leq c_{l-1}(v),~~~\text{for all}~~v \geq v_{l},
\end{equation}
and
\begin{equation}
\label{eq:b3}
c_{l}(v) \leq c_{l+1}(v),~~~\text{for all}~~v \leq v_{l+1},
\end{equation}
then $c_{l}(v) \leq c_{j}(v)$ for $v \in (v_{l}, v_{l+1}),$ for all $j
\in \{0, 1, \ldots, L-1\}.$
\end{lemma}
\proof{
Take $l \in \{0, 1, \ldots, L-1\}$ and a $v \in (v_{l}, v_{l+1}].$ Then
\begin{equation}
c_{l}(v) \leq c_{l+1}(v) \leq c_{l+2}(v) \leq \cdots c_{L-1}(v),
\end{equation}
due to~\eqref{eq:b3} and the fact that $v_{j} < v_{j+1}.$ Also,
by~\eqref{eq:b2},
\begin{equation}
c_{l}(v) \leq c_{l-1}(v) \leq c_{l-2}(v) \leq \cdots c_{0}(v).
\end{equation}
Thus, $c_{l}(v) \leq c_{j}(v)$ for all $j \in \{0, 1, \ldots,
L-1\}.$ \hfill \qed}

We now show, by using Lemma~\ref{lem:opt_cost_prop}, that the
conditions of Lemma~\ref{lem:cond_c} are indeed satisfied and, thus,
it is true that $c_{l}(v) \leq c_{j}(v)$ for $v \in (v_{l}, v_{l+1}),$
for all $j \in \{0, 1, \ldots, L-1\}.$
\begin{enumerate}
\item Note that $c_{0}(v_{1}) = c_{1}(v_{1})$ and $\frac{d c_{0}(v)}{d
    v} > 0$ and $\frac{d c_{1}(v)}{d v} < 0,$ for $v \in (v_{0},
  v_{1}).$ Take $v \in (v_{0}, v_{1}).$ By Taylor series expansion,
  for some $y \in (v, v_{1}),$ we have
      \begin{equation}
        c_{0}(v) = c_{0}(v_{1}) + (v - v_{1})\left.\frac{d c_{0}(v)}{d v}\right|_{v = y} < c_{0}(v_{1}), \label{eq:d1}
      \end{equation}
      where the last inequality follows because $(v - v_{1}) < 0$ and
      $\left.\frac{d c_{0}(v)}{d v}\right|_{v=y} > 0.$ Similarly, for
      some $y \in (v, v_{1}),$ we have
      \begin{equation}
      c_{1}(v) = c_{1}(v_{1}) + (v - v_{1})\left.\frac{d c_{1}(v)}{d v}\right|_{v = y} > c_{0}(v_{1}), \label{eq:d2}
      \end{equation}
      where the last inequality holds as $(v - v_{1}) < 0$ and $\left.\frac{d c_{1}(v)}{d v}\right|_{v=y} < 0.$ Thus, from~\eqref{eq:d1} and~\eqref{eq:d2},
      \begin{equation}
      c_{0}(v) < c_{1}(v), \nonumber
      \end{equation}
      for all $v \in (v_{0}, v_{1}).$

  \item Take $l \in \{1, 2, \ldots, L-2\}.$ Take a $v > v_{l}.$ By mean value theorem, there exists a $z \in (v_{l}, v)$ such that
  \begin{equation}
  \frac{c_{l}(v_{l}) - c_{l}(v)}{c_{l-1}(v_{l}) - c_{l-1}(v)} = \frac{\frac{d c_{l}(z)}{d z}}{\frac{d c_{l-1}(z)}{d z}}. \label{eq:arbit3}
  \end{equation}
  Since, $\frac{d c_{l}(z)}{d z} < \frac{d c_{l-1}(z)}{d z} < 0$ from Lemma~\ref{lem:opt_cost_prop}, we have $\frac{\frac{d c_{l}(z)}{d z}}{\frac{d c_{l-1}(z)}{d z}} > 1.$ Using this reduces~\eqref{eq:arbit3} to
  \begin{equation}
  c_{l}(v_{l}) - c_{l}(v) > c_{l-1}(v_{l}) - c_{l-1}(v), \nonumber
  \end{equation}
  which is nothing but
  \begin{equation}
  c_{l-1}(v) > c_{l}(v_{l}),
  \end{equation}
  since $c_{l}(v_{l}) = c_{l-1}(v_{l}).$

  Similarly, if we take a $v < v_{l+1}$ there exists a $z \in (v_{l},
  v),$ by the mean value theorem, such that
  \begin{equation}
    \frac{c_{l}(v) - c_{l}(v_{l+1})}{c_{l+1}(v) - c_{l+1}(v_{l+1})} = \frac{\frac{d c_{l}(z)}{d z}}{\frac{d c_{l+1}(z)}{d z}}, \nonumber
  \end{equation}
  which reduces to
  \begin{equation}
  c_{l}(v) - c_{l}(v_{l+1}) < c_{l+1}(v) - c_{l+1}(v_{l+1}), \label{eq:arbit4}
  \end{equation}
  by using Lemma~\ref{lem:opt_cost_prop}; but~\eqref{eq:arbit4} is
  nothing but
  \begin{equation}
  c_{l}(v) < c_{l+1}(v),
  \end{equation}
  for all $v < v_{l+1}.$
\end{enumerate}
This proves all the conditions of Lemma~\ref{lem:cond_c}.

\section{Proof of Theorem~\ref{thm:ne_Lq_main}}
\label{app:pf:thm:ne_Lq_main}
Take $v \in \left(v_{l}, v_{l+1}\right],$ then
\begin{align}
F_{l}\left(T_{l}^{\text{NE}}(v)\right) &= \int \mathbbm{1}_{T_{l}^{\text{NE}}(x) \geq T_{l}^{\text{NE}}(v)} q_{l}^{\text{NE}}(v) d\mathcal{G}(v) \nonumber \\
&= \int \mathbbm{1}_{x \leq v} \mathbbm{1}_{v_{l} < v \leq v_{l+1}} d\mathcal{G}(v) = \mathcal{G}(v) - \mathcal{G}(v_{l}), \label{eq:mid_2}
\end{align}
where the second equality follows from Lemma~\ref{lem:ne_Lq_prop}
and~Theorem~\ref{thm:ne_Lq_main_existence}. Differentiating~\eqref{eq:mid_2}
w.r.t. $v,$ we get
\begin{align}
\frac{d\mathcal{G}(v)}{dv} &= \frac{d F_{l}\left(T_{l}^{\text{NE}}(v)\right)}{dv} = \frac{d T_{l}^{\text{NE}}(v)}{d v} \times \left. \frac{d F_{l}(t)}{d t} \right|_{t = T_{l}^{\text{NE}}(v)}. \label{eq:b4}
\end{align}
For optimality, $T_{l}^{\text{NE}}(v)$ must also satisfy the first order condition, namely, $\frac{d c_{v}(l,t)}{d t} = 0.$ This gives
\begin{align}
0 = \left[ N \frac{d F_{l}(t)}{d t} + n_{v}\frac{d g(t)}{d t} \right]_{t = T_{l}^{\text{NE}}(v)}. \label{eq:arbit5}
\end{align}
Substituting~\eqref{eq:b4} in~\eqref{eq:arbit5}, we get
\begin{equation}
\label{eq:b5}
\frac{d g\left(T_{l}^{\text{NE}}(v)\right)}{d v} = -\frac{N}{n_v}\frac{d\mathcal{G}(v)}{dv}. \nonumber
\end{equation}
This gives
\begin{equation}
T_{l}^{\text{NE}}(v) = g^{-1}\left( N \int_{v}^{v_{l+1}} \frac{d\mathcal{G}(x)}{n_{x}} + \alpha\right), \nonumber
\end{equation}
where $\alpha$ is the integration constant that can be shown to equal $0$ using the same line of arguments as in
Appendix~\ref{app:pf:thm:ne_single_p} while deriving
$w^{\text{NE}}(v)$ in explicit form.

Further, since $T_{l}^{\text{NE}}(v)$ is a decreasing function by
Lemma~\ref{lem:ne_Lq_prop}, $T_{l}^{\text{NE}}(v) = 0$ for all $v \geq
v_{l+1}.$ Hence, for $v \geq v_{l+1}$
\begin{align}
  F_{l}\left(T_{l}^{\text{NE}}(v)\right) = F_{l}(0) &= \int \mathbbm{1}_{T_{l}^{\text{NE}}(x) \geq 0} \mathbbm{1}_{v_{l} < v \leq v_{l+1}} d \mathcal{G}(v) \nonumber \\
  &= \int \mathbbm{1}_{v_{l} < v \leq v_{l+1}} d \mathcal{G}(v) \nonumber \\
  &= \mathcal{G}(v_{l+1}) - \mathcal{G}(v_{l}).
\end{align}
Also, note that at Queue $l$ arrival of customer $v = v_l$ is the
earliest. Thus, $F_{l}\left(T_{l}^{\text{NE}}(v_{l})\right) = 0.$ Now,
since $F_{l}\left(T_{l}^{\text{NE}}(\cdot)\right)$ is an increasing
function $F_{l}\left(T_{l}^{\text{NE}}(v_{l})\right) = 0$ for all $v <
v_{l}.$ Thus, any customer $v < v_{l}$ in order to minimize her
waiting cost has to choose $T_{l}^{\text{NE}}(v) =
T_{l}^{\text{NE}}(v_{l}).$ This proves the first part of
Theorem~\ref{thm:ne_Lq_main} which characterizes
$T_{l}^{\text{NE}}(v)$ completely. Notice that, we have also shown
that $T_{l}^{\text{NE}}(v)$ satisfies
\begin{equation}
  F_{l}\left(T_{l}^{\text{NE}}(v)\right) = \left\{ \begin{array}{ll} F_{l}(0) = \mathcal{G}(v_{l+1}) - \mathcal{G}(v_{l}), & \text{for}~~v > v_{l+1} \\
      \mathcal{G}(v) - \mathcal{G}(v_{l}), & \text{for}~~v_{l} \leq v \leq v_{l+1} \\
      0, & \text{for}~~v < v_{l} \end{array}\right. . \label{eq:middle}
\end{equation}

For obtaining the thresholds $\left\{ v_{l} \right\}_{l=1}^{L-1}$ we
look at the optimal queue joining cost.  The optimal joining cost for
Queue $l$ is
\begin{equation}
  c_{l}(v) = N \left[ F_{l}\left( T_{l}^{\text{NE}}(v)\right) +
    \sum_{j=l+1}^{L-1}F_{j}(0) \right] + m_{v} P_l + n_{v} g\left(
    T_{l}^{\text{NE}}(v) \right), \nonumber
\end{equation}
which, by using~\eqref{eq:middle}, reduces to
\begin{equation}
c_{l}(v) = N ( \mathcal{G}(v) - \mathcal{G}(v_{l-1}) ) + N ( 1 - \mathcal{G}(v_{l}) ) + m_{v}P_l + n_{v} N \int_{v}^{v_{l}}. \frac{d\mathcal{G}(x)}{n_{x}}.
\end{equation}
Now, since $v_{l}$ is the point of intersection of $c_{l}(v)$ and
$c_{l-1}(v),$ at $v = v_{l}$ we should have
\begin{align}
  0 &= c_{l}(v_{l}) - c_{l-1}(v_{l}) \nonumber \\
  &= N(1 - \mathcal{G}(v_{l+1})) + m_{v_l}P_l + n_{v_l} N \int_{v_l}^{v_{l}}\frac{d\mathcal{G}(x)}{n_x} - N(1 - \mathcal{G}(v_{l-1})) - m_{v_l}P_{l-1}. \nonumber
\end{align}
This can be re-written as~\eqref{eq:th_def}.

\section{Derivation of the Revenue Function}
\label{app:der:rev_fun}
We can re-write~\eqref{eq:th_def} as
\begin{equation}
\label{eq:rev1}
P_{l} - P_{l-1} = \frac{N}{m_{v_{l}}} \left[ \mathcal{G}(v_{l+1}) - \mathcal{G}(v_{l-1}) - v_{l}\int_{v_{l}}^{v_{l+1}}\frac{d\mathcal{G}(x)}{x} \right],
\end{equation}
for all $l \in \{1, 2, \ldots L-1\}.$ Adding~\eqref{eq:rev1} for all $l = 1$ to $L-1,$ we get
\begin{equation}
\label{eq:rev2}
P_{L-1} = \sum_{l=1}^{L-1} \frac{N}{ m_{v_{l}} }\left[ \mathcal{G}(v_{l+1}) - \mathcal{G}(v_{l-1}) - v_{l}\int_{v_{l}}^{v_{l+1}}\frac{d\mathcal{G}(x)}{x} \right].
\end{equation}
We use $\mathbf{P}$ to denote the price vector $(P_{1}, \ldots, P_{L-1}).$ Revenue function can be expanded as
\begin{align}
\frac{1}{N}R(\mathbf{P}) &= \sum_{l=1}^{L-1}\left[ \mathcal{G}(v_{l+1}) - \mathcal{G}(v_{l}) \right] P_{l} = P_{L-1} - \sum_{l=1}^{L-1} \left(P_{l} - P_{l-1}\right) \mathcal{G}(v_{l}).
\end{align}
Substituting~\eqref{eq:rev1} and~\eqref{eq:rev2}, we obtain
\begin{multline}
  \frac{1}{N}R(\mathbf{P}) =  \sum_{l=1}^{L-1}\! \frac{N}{ m_{v_{l}} } \left[ \mathcal{G}(v_{l+1}) - \mathcal{G}(v_{l-1}) - v_{l}\int_{v_{l}}^{v_{l+1}}\frac{d\mathcal{G}(x)}{x} \right]\\
  - \sum_{l=1}^{L-1} \mathcal{G}(v_{l})\frac{N}{ m_{v_{l}} } \left[ \mathcal{G}(v_{l+1}) -
    \mathcal{G}(v_{l-1}) - v_{l}\int_{v_{l}}^{v_{l+1}}\frac{d\mathcal{G}(x)}{x} \right],
\end{multline}
which can be written as
\begin{equation}
  \frac{1}{N}R(\mathbf{P}) = \sum_{l=1}^{L-1} \left( 1 - \mathcal{G}(v_{l}) \right)\frac{N}{m_{v_{l}}}
  \times \left[\mathcal{G}(v_{l+1})  - \mathcal{G}(v_{l-1}) - v_{l}\int_{v_{l}}^{v_{l+1}}\frac{d\mathcal{G}(x)}{x} \right].
\end{equation}
This can be simplified as
\begin{multline}
  \frac{1}{N}R(\mathbf{P}) = \sum_{l=1}^{L-1}\frac{N}{m_{v_{l}}}\left[\mathcal{G}(v_{l+1}) - v_{l}\int_{v_{l}}^{v_{l+1}}\frac{d\mathcal{G}(x)}{x} \right] \left( 1 - \mathcal{G}(v_{l}) \right) \\
  - \sum_{l=1}^{L-1} \left( 1 - \mathcal{G}(v_{l})
  \right)\frac{N}{m_{v_{l}}} \mathcal{G}(v_{l-1}).
\end{multline}
Note that the first term in the second sum is $0.$ Hence, by
increasing index of the second summation this can be re-written as
\begin{multline}
  \frac{1}{N}R(\mathbf{P})= \sum_{l=1}^{L-1} \frac{N}{m_{v_{l}}} \left[ \mathcal{G}(v_{l+1})  - v_{l} \int_{v_{l}}^{v_{l+1}} \frac{d\mathcal{G}(x)}{x} \right] \left( 1 - \mathcal{G}(v_{l}) \right) \\
  - \sum_{l=1}^{L-2} \left( 1 - \mathcal{G}(v_{l+1}) \right)\frac{N}{m_{v_{l+1}}} \mathcal{G}(v_{l}).
\end{multline}
This is noting but $\frac{1}{N}R(\mathbf{P}) = \sum_{l=1}^{L-1}
u(v_{l}, v_{l+1}).$

\bibliographystyle{abbrv}


\end{document}